\documentclass[twocolumn,amsmath,amssymb,showpacs,aps,pra]{revtex4-1}

\usepackage{amsmath}
\usepackage{amssymb}
\usepackage{graphicx}
\usepackage[toc,page]{appendix}
\usepackage{color}

\newcommand{\eqa}{\begin{eqnarray}}
\newcommand{\eeqa}{\end{eqnarray}}
\newcommand{\beq}{\begin{equation}}
\newcommand{\eeq}{\end{equation}}

\newcommand{\eps}{\epsilon}
\newcommand{\benumerate}{\begin{enumerate}}
\newcommand{\eenumerate}{\end{enumerate}}

\newcommand{\bitemize}{\begin{itemize}}

\newcommand{\eitemize}{\end{itemize}}

\newcommand{\dertot}[2]{\frac{d #1}{d #2}}

\begin{document}

\title{Breaking mechanism from a vacuum point\\ in the defocusing nonlinear Schr\"odinger equation}

\author{Antonio Moro$^1$}
\email{antonio.moro@northumbria.ac.uk}
\author{Stefano Trillo$^2$}
\affiliation{(1) Department of Mathematics and Information Sciences, Northumbria University, Newcastle upon Tyne, UK\\
(2) Dipartimento di Ingegneria, Universit\`a di Ferrara, Italy}

\date{\today} 

\begin{abstract}
We study the wave breaking mechanism for the weakly dispersive defocusing nonlinear Schr\"odinger (NLS) equation with a constant phase dark initial datum that contains a vacuum point at the origin. We prove by means of the exact solution to the initial value problem that, in the dispersionless limit, the vacuum point is preserved by the dynamics until breaking occurs at a finite critical time. In particular, both Riemann invariants experience a simultaneous breaking at the origin. Although the initial vacuum point is no longer preserved in the presence of a finite dispersion, the critical behavior manifests itself through an abrupt transition occurring around the breaking time.
\end{abstract}

\maketitle

\section{Introduction}
The nonlinear Schr\"odinger (NLS) equation
\begin{equation}
\label{NLS}
i \eps  \psi_{t} + \frac{\eps^{2}}{2} \psi_{xx} + \sigma  |\psi|^{2} \psi =0, \qquad \sigma =\pm 1
\end{equation}
where $\eps$ is a positive constant parameter, in its focusing ($\sigma = 1$) and defocusing ($\sigma = -1$) versions has a universal character as it arises in the Hamiltonian description of envelope waves for a general class nonlinear model equations~\cite{CAL}. For this reason the NLS equation naturally appears in a variety of physical contexts such as nonlinear optics \cite{agrawalbook}, water waves \cite{osbornebook} and superfluidity \cite{GPeq,RB}, including atom condensates \cite{DGPS99} and quantum fluid behavior of light \cite{CC13}. Moreover, the remarkable mathematical structure of the NLS equation also explains its relevance for physical applications as well as a mathematical object itself. Similarly to the celebrated Korteweg-de Vries (KdV) equation the NLS equation has been shown to be an example of integrable nonlinear PDE admitting infinitely many conservation laws and soliton solutions which can be constructed via the Inverse Scattering Transform method~\cite{AS,NMPZ,ZS}. Since the discovery of its integrability, the NLS equation~(\ref{NLS}) and its generalizations have been subject matter of intensive studies covering both analytical~\cite{APT}  and algebro-geometric aspects~\cite{BBEIM}. 

An important feature, common to many nonlinear dispersive PDEs, is the critical behavior connected to the appearance, in the dispersionless limit,
of one or more breaking points typically due to gradient catastrophies. Such gradient catastrophies are regularized by the finite dispersion through the emission of wavetrains, so-called dispersive shock waves (also known as collisionless shock waves from the early literature on plasma \cite{GP} or undular bores \cite{UB} mainly in hydrodynamics). Attention on such phenomenon was originally drawn in the framework of the KdV equation \cite{ZK65}, for which the analytical construction of dispersive shocks was pioneered by Gurevich
 and Pitaevskii~\cite{GP} via the Whitham averaging method \cite{Whitham65}, and improved afterwards on a more rigorous basis~\cite{FFM80,LL83,V85,K,Tian94}.
These approaches have been successfully extended to the defocusing NLS equation \cite{GK87,P87,GKE92,ElKrylov95,EGGK95,TY99,JLM1,JLM2,WFM99,Kodama99,Kam04,Hoefer06,Fratax08}, and employed to effectively describe recent experiments in condensates of dilute atom gases \cite{Dutton01,Hoefer06,Chang08,Mep09} and nonlinear optics \cite{RG89,Wan07,Gofra07,Fleischer07,CFPRT09}.

In the recent years a special attention has been devoted to the study of gradient catastrophe phenomena in the weakly dispersive regime and to the underlying mechanisms. An important advance in this direction was the notion of universality introduced in 2006 by Dubrovin to describe the wave-breaking associated with Hamiltonian systems. The {\em universality conjecture} is concerned with the asymptotic description of the critical behavior for {\it generic initial data} such that the dispersionless limit of the PDE is either strictly hyperbolic (e.g. defocusing NLS) or elliptic (e.g. focusing NLS) near the critical point of gradient catastrophe \cite{D06,D08,CG09,DGK09,BT12,DGKM}. Such an approach has also been shown to be effective for other classes of non-Hamiltonian PDEs~\cite{ALM,DE}.
There are, however, certain cases where the initial data (wavepackets) are non generic, demanding for a specific analysis of the breaking mechanism. Our aim, in this paper, is to report such analysis for the case of the defocusing NLS subject to an initial condition $\psi_0=\psi(x,0)$ possessing a {\it vacuum point} (a point where the field strictly vanishes). This case was indeed recently considered in Ref.~\cite{CFPRT09} for $\psi_0=\tanh (x)$, i.e. a dark amplitude with step-like phase. It was numerically shown that the wave-breaking taking place through an apparent singularity developing in the vacuum state is resolved through the generation of a dispersive shock wave constituted by an expanding oscillating fan centered around a still dark (black) soliton (hence the vacuum state turns out to be preserved by the dynamics across the critical region). Such scenario was also observed experimentally and shown to be robust against the nonlocal deformation of the NLS model \cite{CFPRT09,Armaroli09}, and to bear close similarity with the breaking mechanism in the periodic case $\psi_0=\cos (x)$, where an array of breaking points develop through multiple four-wave mixing \cite{Trillo10}.
Nevertheless, in the vacuum point the dispersionless NLS equation looses its strict hyperbolicity and the mechanism of breaking differs from the case of generic initial data  making the universality conjecture not applicable. In order to investigate the breaking mechanism for an initial state with a vacuum point and to be able to describe the wave-breaking analytically, we focus in this paper on the study of the NLS equation (\ref{NLS}) with $\sigma=-1$ and $\eps \ll 1$, subject to the dark constant phase initial condition 
\begin{equation}
\label{initdatum}
\psi_0 = \left | \tanh x \right | \; e^{i \theta_{0}}
\end{equation}
where $\theta_{0}$ is a real constant. Hence,  we compare the dispersionless dynamics in the vicinity of the gradient catastrophe with the numerical simulations for the full NLS equation. The main difference with the case studied in~\cite{CFPRT09} is that in this case the vacuum is not preserved by the evolution as a consequence of the absence of the phase jump. Nevertheless, in the dispersionless limit, the qualitative behavior is similar until the time of gradient catastrophe is reached. We show that in the dispersionless regime the vacuum point is a critical point associated with a gradient catastrophe occurring simultaneously for the two Riemann invariants. A detailed analytical description of the wave breaking mechanism is provided, and the deviation due to the finite value of dispersion in the NLS equation are analyzed in details. 
 
\section{Zero dispersion limit}
The zero dispersion regime for the NLS equation is concerned with the study of fast oscillating solutions $\psi(x,t;\eps)$ in the limit $\eps \to 0$. For this purpose it is convenient to introduce the nonlinear (Madelung) transformation of the form
\begin{equation}
\psi(x,t;\eps) = \sqrt{u(x,t)} \exp \left( \frac{i}{\eps} \int^{x} v(x',t) \; dx' \right ).
\end{equation}
In terms of the variables $u$ and $v$, having the meaning of an equivalent fluid density (or height) and velocity,  the NLS equation~\eqref{NLS} takes the so-called hydrodynamic form
\begin{gather}
\label{hydro_NLS}
\begin{aligned}
&u_{t} + (u v)_{x} = 0, \\
&v_{t} +v v_{x} + u_{x}- \frac{\eps^{2}}{4} \left(\frac{u_{xx}}{u} - \frac{u_{x}^{2}}{2u^{2}} \right)_{x} = 0.
\end{aligned}
\end{gather}
At leading order in $\eps$, neglecting  the so-called {\it quantum pressure} term of order $O(\eps^{2})$, Eqs.~(\ref{hydro_NLS}) give the dispersionless limit which turn out to be equivalent to the following well known system that rules 1D shallow water waves - henceforth shallow water equations (SWE) - or equivalently isentropic gas dynamics with pressure proportional to square density, 
\begin{gather}
\label{dNLS}
\begin{aligned}
&u_{t} + (u v)_{x} = 0, \\
&v_{t} +v v_{x} + u_{x}= 0.
\end{aligned}
\end{gather}
Higher-order corrections to the solutions of Eqs. (\ref{dNLS}) can be formally constructed by means of power expansion of $u,v$, which, however, will not be considered further in this paper. Conversely we will directly compare the breaking dynamics entailed by the SWE (\ref{dNLS}) with the smooth evolution according to the full NLS equation.

\begin{figure}[h!]
\centering
\includegraphics[width=9cm]{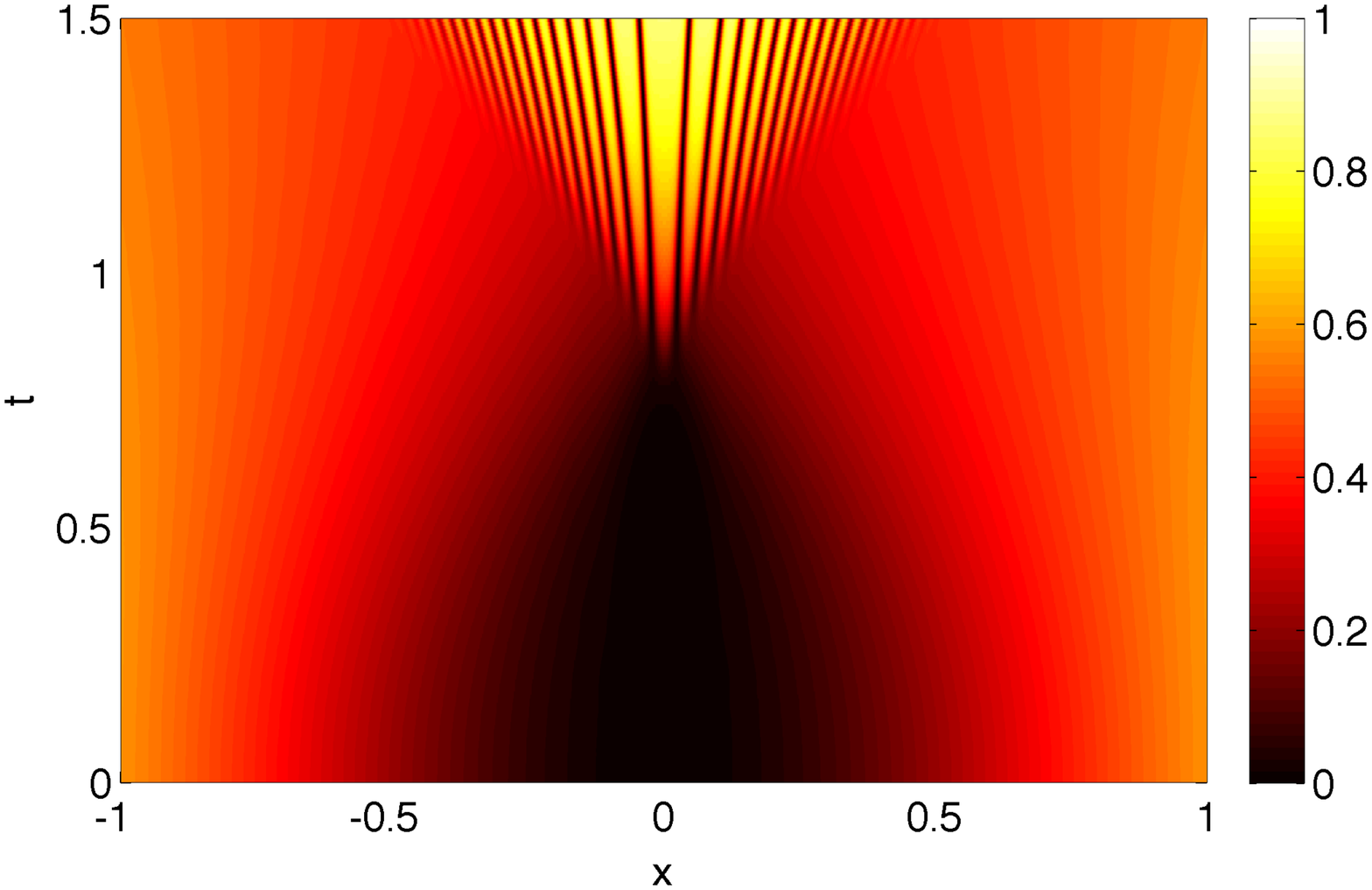}
\caption{(Color online) Color level plot of the evolution of $|\psi|^2$ from the input $\psi_0=\sqrt{(\tanh x)^2}$ according to the defocusing ($\sigma=-1$) NLS equation, with $\varepsilon=0.005$.
} 
\label{fig1} \end{figure}

\begin{figure}[h!] 
\centering
\includegraphics[width=8cm]{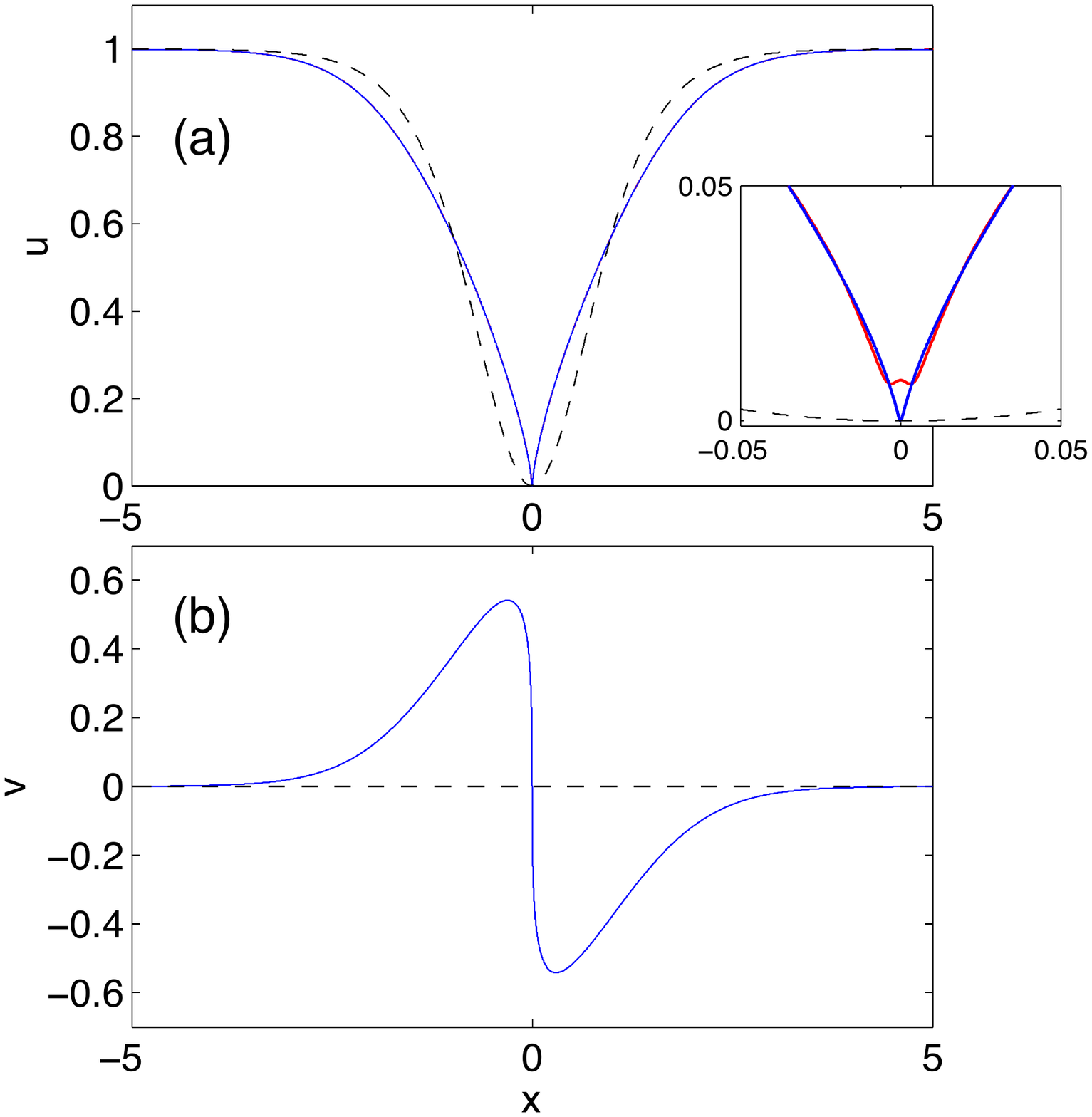}
\caption{(Color online) Snapshots from SWE [Eqs.~(\ref{dNLS})] at $t=0.78$ (close to breaking): (a) density $u$; (b) velocity $v$. The inset shows, in a zoom close to the origin, the deviation in terms of density $u$ of the NLS dynamics (red curve; obtained from Eq. (\ref{NLS}) with $\varepsilon=5 \times 10^{-4}$) from the SWE (blue curve).
The black dashed lines stand for the initial condition.} 
\label{fig2} \end{figure}

Figure~\ref{fig1} shows the evolution of the dark initial datum (\ref{initdatum}) according to the Eq.~(\ref{NLS}) computed numerically. A dispersive shock that opens up in a characteristic fan turns out to resolve the singularity that occurs in the range $t=0.75-0.78$ around the origin ($x=0$), where the initial dark dip appears to focus before breaking. The numerical integration of the SWE [see Fig. \ref{fig2}] shows a wave-breaking scenario compatible with a breaking in the origin (or at least around it) at $t \simeq 0.78$, where the density $u$ still exhibits a vacuum point and the velocity $v$ develops a steep gradient and takes positive and negative values for $x<0$ and $x>0$, respectively. However, as shown in the inset in Fig. \ref{fig2}(a), the initial vacuum point, which is preserved by the SWE dynamics, is not preserved in the NLS dynamics. These observations call for a deeper analysis of the problem. The advantage of considering the SWE~(\ref{dNLS}) is that it provides an accurate description of the system before the breaking time~\cite{JLM1,JLM2} and it can be effectively treated via the classical hodograph method. Indeed, for sufficiently small values of the dispersive parameter $\eps$, the local minimum evolving from the initial vacuum state stays close to zero until it turns into a local maximum only in proximity of the critical time.


The hydrodynamic type of system~(\ref{dNLS}) is linearized via the hodograph transform $(x,t) \leftrightarrow (u,v)$, that is obtained by interchanging the role of dependent and independent variables via the following system
\begin{equation}
\label{hod}
v t + f_{u} = x, \qquad \qquad  u t + f_{v} = 0,
\end{equation}
where the function $f(u,v)$ is a solution to the Tricomi-type equation
\begin{equation}
\label{tricomi}
f_{vv} - u f_{uu} = 0.
\end{equation}
Given a solution to the equation~(\ref{tricomi}) the corresponding solution to the SWE~(\ref{dNLS})  is locally given by the functions $u =u(x,t)$ and $v=v(x,t)$ obtained by inversion of the hodograph equations~(\ref{hod}). The general procedure for solving the initial value problem for the system~(\ref{dNLS}) has been discussed in~\cite{TY99}.
Nevertheless, an explicit analytical description of solutions and the exact computation of critical values is possible in a limited number of cases, which are however of great interest for understanding various wave breaking mechanisms, observed also experimentally.

For the study of wave breaking it is convenient to introduce the Riemann invariants
\begin{align}
\label{RInvariants}
\xi = v + 2 \sqrt{u}, \qquad \eta = v - 2 \sqrt{u},
\end{align}
such that the SWE~(\ref{dNLS}) takes the diagonal form
\begin{equation}
\label{dNLS_riemann}
\xi_{t} + \lambda \xi_{x} = 0, \qquad  \eta_{t} + \mu \eta_{x} = 0,
\end{equation}
where $\lambda(\xi,\eta)$ and $\mu(\xi,\eta)$ are the characteristic speeds
\begin{align*}
\lambda  = \frac{3 \xi +\eta}{4}, \qquad \qquad \mu  = \frac{\xi + 3 \eta}{4}.
\end{align*}
In terms of Riemann invariants, Eqs.~(\ref{hod}) read as follows
\begin{equation}
\label{hodR}
x -\lambda t = w_{\xi}, \qquad \qquad  x -\mu t = w_{\eta},
\end{equation}
and the Tricomi type equation~(\ref{tricomi}) is replaced by the Euler-Poisson-Darboux (EPD) equation for the function $w(\xi,\eta)$
\begin{equation}
\label{EPD}
w_{\xi\eta} = \frac{1}{2(\xi-\eta)} \left(w_{\xi} - w_{\eta} \right).
\end{equation}
Differentiating the hodograph equations~(\ref{hodR}) by $x$ and $t$ and solving w.r.t. $\xi_{x}$, $\xi_{t}$, $\eta_{x}$, $\eta_{t}$, we get
\begin{gather}
\label{rbreak}
\begin{aligned}
\xi_{x} = \frac{2 (\eta - \xi)}{2 (\eta -\xi) w_{\xi\xi} + 3 (w_{\xi} - w_{\eta})} \\
\eta_{x} = \frac{2 (\eta - \xi)}{2 (\eta -\xi) w_{\eta\eta} + 3 (w_{\xi} - w_{\eta})}.
\end{aligned}
\end{gather}
The solution $(\xi(x,t), \eta(x.t))$ is said to break, i.e. it develops a gradient catastrophe singularity, if there exists a critical point $(x_{c}, t_{c})$ such that $\xi_{x}$ and $\eta_{x}$ are bounded for any $t \in [0,t_{c})$ and $|\xi_{x}(x_{c}, t_{c})| = \infty$ or $|\eta_{x}(x_{c}, t_{c})| = \infty$. Hence, the gradient catastrophe is associated with the minimum time such that at least one of Riemann invariants breaks at a certain $x = x_{c}$.  From expressions~(\ref{rbreak}) it follows that a necessary condition for the breaking is given by the vanishing condition of  one of denominators in~Eq. (\ref{rbreak}). If Riemann invariants do not break simultaneously at the same point $x$, provided the system is strictly hyperbolic, the critical point ($x_{c},t_{c}$) is said to be {\it generic}~\cite{D06}. 
Necessary condition for simultaneous breaking is that both denominators in Eqs.~(\ref{rbreak}) vanish at one and the same point. In this case the critical point is said to be {\em non-generic}.

For a generic critical point such that, say, only $\xi_{x}$ blows up, we have
\begin{equation}
\label{xibreak}
w_{\xi\xi} = \frac{3}{2 (\xi -\eta)} \left(w_{\xi} -w_{\eta} \right).
\end{equation}
We note also that, if $w_{\xi\xi} \neq w_{\eta\eta}$ at ($x_{c}, t_{c}$) and Riemann invariants are bounded, then $\eta_{x}$ is also bounded. From  Eqs.~(\ref{hodR}) we have that the critical time $t=t_c$ is given by
\begin{equation}
\label{tcrit}
t_c = - \frac{4}{3} w_{\xi \xi}(\xi_{c},\eta_{c}),
\end{equation}
where the local minimum condition
\begin{equation}
\label{mintime}
w_{\xi\xi\xi} (\xi_c,\eta_c) = 0,   \qquad w_{\xi\xi\eta}(\xi_c,\eta_c)  = 0,
\end{equation}
has to be satisfied and the pair $(\xi_{c}, \eta_{c})$ simultaneously solves the breaking condition~(\ref{xibreak}). Let us also observe that only two of the three conditions in Eq.~(\ref{xibreak}) and Eq.~(\ref{mintime}) are independent, for instance
\begin{align*}
&w_{\xi\xi} - \frac{3}{2 (\xi -\eta)} \left(w_{\xi} -w_{\eta} \right) =0, &w_{\xi\xi\xi} = 0.
\end{align*}
Hence, given $(\xi_{c},\eta_{c})$ to be roots of the above system, the critical point ($x_{c},t_{c}$) is obtained from 
Eqs.~(\ref{hodR})
evaluated at ($\xi_{c}, \eta_{c}$), provided the following condition on the Hessian is verified 
\begin{equation*}
{\cal H} = \left |
\begin{array}{cc}
t^{c}_{\xi\xi}& t^{c}_{\xi\eta}\\
\\
t^{c}_{\xi\eta} & t^{c}_{\eta\eta}
\end{array}
\right | > 0.
\end{equation*}
The above condition guarantees that the stationary point~(\ref{tcrit}) is a local minimum as necessary to identify the time when the first gradient catastrophe occurs.
According with the universality conjecture~\cite{D08} (see also~\cite{DGK09,DGKM}) for a generic critical point of gradient catastrophe, critical values characterize the local asymptotic behavior of the perturbed dispersive system.  In particular, the normal form of the  solution to a dispersionless hyperbolic system of Hamiltonian PDEs near the critical point is given by a Whitney singularity of Type 2. In the weak dispersive regime, the leading asymptotic behavior is given by a Painlev\'e trascendent obtained as a particular solution to the PI2 equation. In this case all parameters such as amplitude and scaling factors are completely specified by the critical values for the dispersionless system. 

In the following, we show that the initial datum~(\ref{initdatum}) that contains a vacuum point develops a non-generic gradient catastrophe. Therefore, the universality conjecture does not apply and consequently a detailed description of the breaking mechanism requires a separate analysis. Moreover, previous analysis on the general role of a vacuum point in models of gas or fluid dynamics do not help to unveil the details of the breaking mechanism \cite{Smoller80,Yang06}.  

\section{Solution for a constant phase initial datum}
The hodograph method allows to construct, at least locally, any solution to the SWE~(\ref{dNLS}), with the exception of a neighbourhood of stationary points for Riemann invariants.

Solving the initial value problem requires the knowledge of a suitable solution to the linear equation~(\ref{tricomi}) such that the functions $u(x,t)$ and $v(x,t)$ obtained from the inversion of the system~(\ref{hod}) match the initial condition at the time $t=0$.  In other words the solution to the initial value problem consists of the construction of a map from the space of initial data to the space of solutions to Eq.~(\ref{tricomi}) or, equivalently, the EPD equation~(\ref{EPD}). This problem has been solved for defocusing dispersionless NLS equation in Ref.~\cite{TY99}, while it  remains open in the case of a general system of hydrodynamic type. In the following, we present an alternative approach that applies to a more restricted class of initial data but that can be straightforwardly extended to more general two-component systems of hydrodynamic type. 

Let us consider the family of initial data of the form
\begin{align}
\label{gen_init}
&u(x, 0) = u_{0}(x) & v(x,0) = 0.
\end{align}
For the sake of simplicity, $u_{0}$ is assumed to be a negative hump given by an even,  continuous and differentiable function centered at $x=0$.
As the the change of variables~(\ref{RInvariants}), reducing SWE~(\ref{dNLS}) to the diagonal form~(\ref{dNLS_riemann}) is not globally differentiable for the initial value problem under consideration, we prefer to present the solution to the initial value problem~(\ref{gen_init}) using the natural variables~($u$,$v$) and then introduce Riemann invariants for the local analysis around the breaking point.

Let us observe that the function
\begin{equation}
\label{fg}
f(u,v) = u \int_{-1}^{1} g(v+2 \mu \sqrt{u}) \sqrt{1-\mu^{2}} \; d\mu,
\end{equation}
where $g$ is an arbitrary function of one variable such that the integral is well defined, gives a family of solutions to the Tricomi equation~(\ref{tricomi}). This can be straightforwardly proven by direct substitution and integration by parts. 
As the Tricomi equation~(\ref{tricomi}) is related to the EPD equation~(\ref{EPD}) via the transformation~(\ref{RInvariants}), solutions of the form~(\ref{fg}) to Eq.~(\ref{tricomi}) can be found as a particular case of the general solution to the EPD equation discussed in Ref.~\cite{ER70} (up to a suitable change of variables). Alternatively, after the transformation~(\ref{RInvariants}), formula~(\ref{fg}) can be obtained from a reduction of the formula~(2.5) in~Ref.~\cite{Tian94}.

Observing that
\begin{eqnarray*}
f_{u}(u,v) &=\frac{1}{2} \int_{-1}^{1} g(v+2 \mu \sqrt{u}) \frac{d\mu}{\sqrt{1-\mu^{2}}}, \\
f_{v}(u,v) &= u  \int_{-1}^{1} g'(v+2 \mu \sqrt{u}) \sqrt{1-\mu^{2}} \; d\mu,
\end{eqnarray*}
the hodograph Eqs.~(\ref{hod}) evaluated for the initial condition~(\ref{gen_init}) give the following integral equation for the function $g$
\begin{subequations}
\begin{align}
\label{hod21}
&\frac{1}{2} \int_{-1}^{1} g(2 \mu \sqrt{u_{0}}) \frac{d\mu}{\sqrt{1-\mu^{2}}} = x \\
\label{hod22}
& u_0 \int_{-1}^{1} g'(2 \mu \sqrt{u_{0}}) \sqrt{1-\mu^{2}} \; d\mu =0.
\end{align}
\end{subequations}
Looking for solutions of the form~(\ref{fg}) such that $g$ is an even function of its argument, i.e. $g(r) = g(-r)$, Eq.~(\ref{hod22}) turns out to be identically satisfied. Hence, restricting our analysis to the positive half-line, Eq.~(\ref{hod21}) can be equivalently written in the following form
\begin{equation}
\label{abel}
\int_{0}^{s}  \frac{h(\tau)}{\sqrt{s-\tau}} \; d\tau = \lambda(s),
\end{equation} 
where 
\[
h(\tau) = \frac{g(\sqrt{\tau})}{2 \sqrt{\tau}};~\tau = 4 u_{0} \mu^{2}; ~ \lambda(s) = u_{0}^{-1} \left( \frac{s}{4} \right);~ s =4 u_{0}.
\]
Equation~(\ref{abel}) can be solved with respect to the unknown function $h(\tau)$ via the Abel transform. The solution
\begin{align*}
h(\tau) &= \frac{1}{\pi} \dertot{~}{\tau} \int_{0}^{\tau} \frac{\lambda(s)}{\sqrt{\tau - s}} \; ds = \frac{1}{\pi} \int_{0}^{\tau} \frac{\lambda'(s)}{\sqrt{\tau-s}} \; ds + \frac{\lambda(0)}{\pi \sqrt{\tau}},
\end{align*}
gives
\begin{gather}
\begin{aligned}
\label{gform}
g(r)& = 2 \dertot{~}{r}  \int_{0}^{r} \frac{s \lambda(s^{2})}{\sqrt{r^{2} - s^{2}}} \; ds\\  
&= \frac{2 r}{\pi} \left [ \int_{0}^{r}  \frac{2 s \lambda'(s^{2})}{\sqrt{r^{2}- s^{2}}}  \; ds + \frac{\lambda(0)}{r}\right ], \qquad r>0 \;.
\end{aligned}
\end{gather}
Hodograph equations~(\ref{hod}) together with the formulae~(\ref{fg}) and~(\ref{gform}) provide the solution to the present class of initial value problems for the zero dispersion limit of the NLS equation. \\
Let us now consider the particular initial datum~(\ref{initdatum}) which in the hydrodynamic variables reads as
\begin{equation*}
u(x,0) = \tanh^{2}( x ) \qquad v(x,0)=0.
\end{equation*}
It belongs to the class~(\ref{gen_init}) and, as mentioned above, it contains a vacuum point at the origin ($x=0$).
In this case, the expression~(\ref{gform}) takes the following simple form
\begin{equation}
\label{gex}
g(r) = \frac{|r|}{\sqrt{4 - r^{2}}} \qquad -2 < r < 2.
\end{equation}
Introducing the variable $r = v + 2 \mu \sqrt{u}$ in Eq.~(\ref{fg}), away from the vacuum, $f(u,v)$ is expressed in terms of Riemann invariants $(\xi, \eta)$
by defining the function $F(\xi,\eta) \equiv f(u(\xi,\eta),v(\xi,\eta))$
\begin{widetext}
\begin{equation} \label{FR}
F(\xi,\eta)=\frac{1}{4} \int_{\eta}^{\xi} g(r) \sqrt{(\xi-r) (r-\eta)} \; dr =\frac{1}{4} \int_{0}^{\xi} \frac{r \sqrt{(\xi-r) (r-\eta)}}{\sqrt{4-r^{2}}}~dr + \frac{1}{4} \int_{0}^{\eta} \frac{r \sqrt{(\xi-r) (r-\eta)}}{\sqrt{4-r^{2}}}~dr 
\end{equation}
\end{widetext}

\begin{figure}[h!]
\includegraphics[width=9cm]{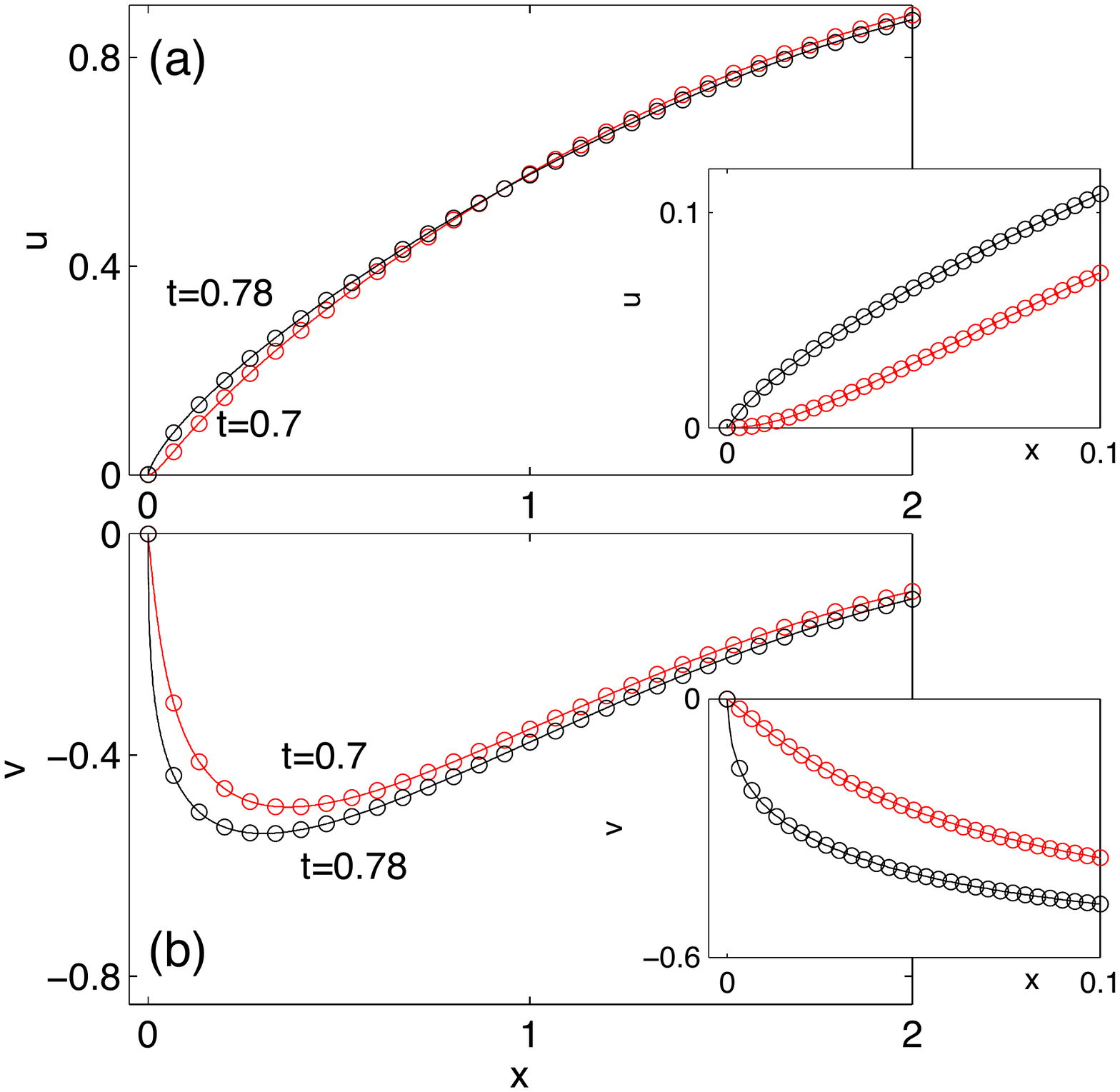}
\caption{(Color online) Comparison of analytic (open circles) and numerical (solid line) solutions of dispersionless limit [SWE, Eqs. (\ref{dNLS})] sampled at the times $t=0.7, 0.78$.
The insets show a zoom of the region close to the origin. The maximum deviation between the two sets of solutions turns out to be $\sim 5 \times 10^{-5}$.
} 
\label{figexnum}
\end{figure} 
The elliptic integrals in the RHS of Eq.~(\ref{FR}) can be evaluated as a combination of complete and incomplete elliptic integrals and Jacobian elliptic functions.
We report such calculation explicitly in Appendix~\ref{app1}. 
Since the function $u=u(x,t)$ and $v=v(x,t)$ are obtained by inversion of the hodograph formulas~(\ref{hod}), which is not straightforward in a neighbourhood of the critical point, as a further proof of validity of our calculations, we compare the results obtained via the analytic formulas with the direct numerical integration of the SWE obtained via a finite-difference method (the latter integration is not critical since is performed up to the breaking point where solutions are smooth).
Figure \ref{figexnum} shows a direct comparison between the analytic solution obtained via Eqs.~(\ref{hod}) by using a standard Newton's method, and the numerical solution of the SWE.
The agreement is excellent and apparently, the dependent variable $v$ develops a gradient catastrophe at $x=0$, with the critical time being estimated to be $t_{c} \simeq 0.78$.
In the next section, we prove that the initial vacuum point $x =0$ is preserved in the dispersionless limit [SWE, Eqs.~(\ref{dNLS})] and show that it is a critical point of gradient catastrophe for which we compute the breaking time exactly.

\section{Vacuum point and critical time for breaking}
We start by studying the invariance properties associated with the vacuum point.
Let us consider a smooth solution $u(x,t)$, $v(x,t)$ to SWE~(\ref{dNLS}) such that the initial datum $u(x,0)$, $v(x,0)$ contains a vacuum point at $x=x_{0}$, where the derivative also vanishes, that is
\[
u(x_0,0) =0, \; v(x_0,0) =0,  \; u_x(x_0,0) =0.
\]
We observe that under the assumptions above we have 
\begin{equation}
\label{vp}
u(x_0,t) =0, \; v(x_0,t) =0,  \; u_x(x_0,t) =0,
\end{equation}
for any time $0 \leq t <t_{c}$. 

Indeed, let us first differentiate the functions $u(x,t)$, $v(x,t)$ and $u_{x}(x,t)$ along a generic curve $(t, x(t))$ 
\begin{gather}
\label{stat}
\begin{aligned}
\dertot{u}{t} &=u_{x}(x(t),t) x'(t) + u_{t}(x(t), t),  \\
\dertot{v}{t} &=v_{x}(x(t),t) x'(t) + v_{t}(x(t), t), \\
\dertot{u_{x}}{t} &=u_{xx}(x(t),t) x'(t) + u_{xt}(x(t), t).
\end{aligned}
\end{gather}
Using the fact that $u$ and $v$ satisfy SWE~(\ref{dNLS})  and restricting to the particular curve $(t,x_{0})$ expressions~(\ref{stat}) gives the following non-autonomous dynamical system
\begin{gather}
\label{stat2}
\begin{aligned}
\dot{U} &= -  \alpha U -  W  V \\
\dot{V} &= -  \alpha V - W \\
\dot{W} &= - \gamma U - \beta V - 2 \alpha W,
\end{aligned}
\end{gather}
where $U(t) = u(x_{0},t)$, $V(t) = v(x_{0},t)$, $W(t) = u_{x}(x_{0},t)$, parametrised by the three independent functions of time $\alpha(t) = v_{x}(x_{0},t)$, $\beta(t) = u_{xx}(x_{0},t)$, $\gamma(t) = v_{xx}(x_{0},t)$.

Clearly the vacuum point~(\ref{vp}) is a stationary solution to the above dynamical system for any $t <t_{c}$, where $t_{c}$ is the time of gradient catastrophe, such that the functions  
$\alpha(t)$, $\beta(t)$, $\gamma(t)$ are bounded. We do not enter into the stability analysis for the dynamical system~(\ref{stat2}) as this lies out of the scope of the present work.

\begin{figure}[htb!] 
\centering
\includegraphics[width=8cm]{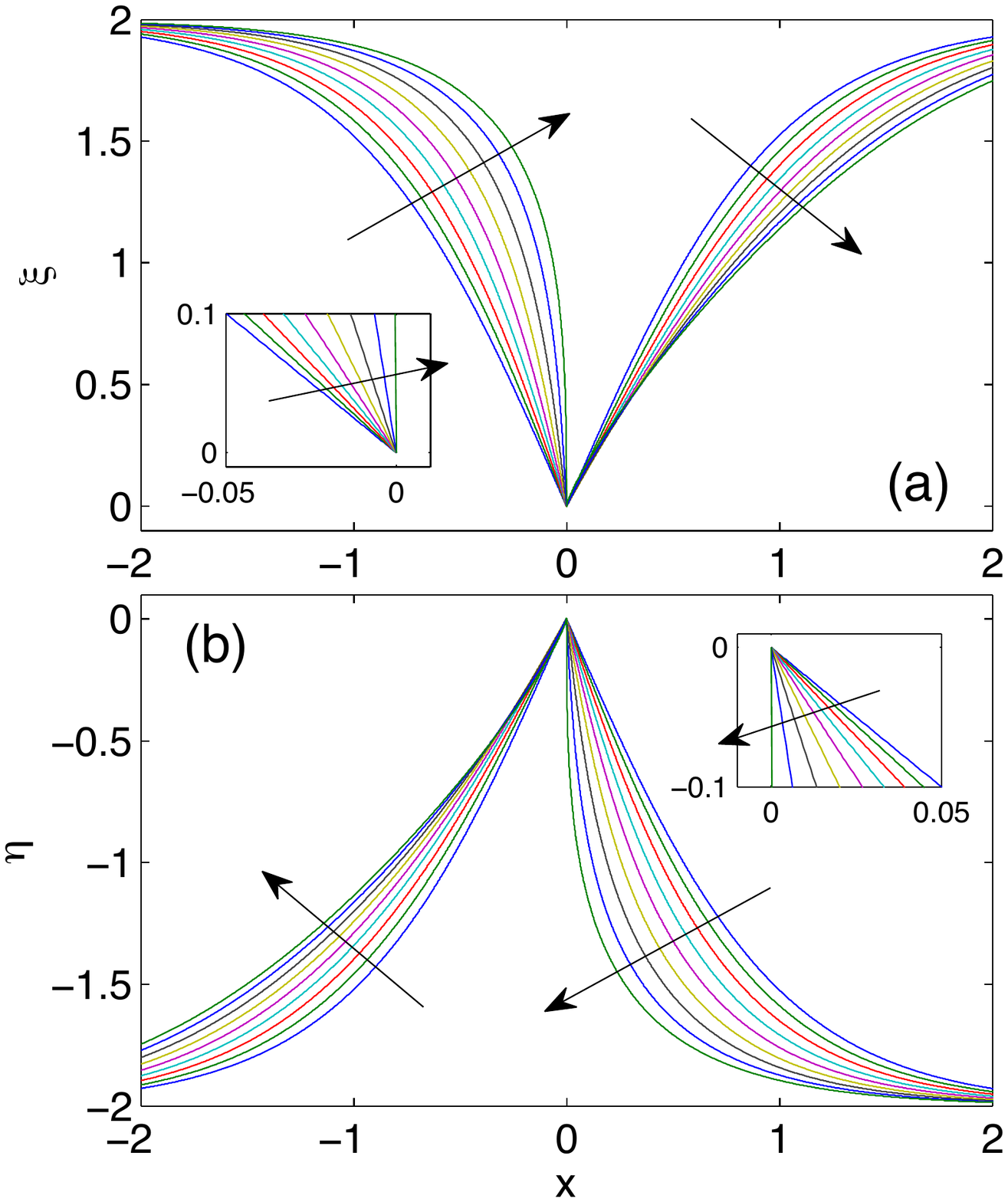}
\caption{(Color online) Snapshots of Riemann invariants vs. $x$ obtained by means of numerical integration of the dispersionless equations [Eqs. (\ref{dNLS})].
The insets show a zoom close to the origin on the semi-axis where the relative Riemann invariant breaks.
The snapshots are taken at time $t=0,0.1,0.2,0.3,0.4,0.5,0.6,0.7,0.784$, and the arrows give the direction in which time $t$ grows.
} 
\label{SWEriemann}\end{figure}

We now proceed proving that the vacuum point is a point of gradient catastrophe and compute the critical values.
This will be done via an asymptotic evaluation of the the formula ~(\ref{fg})  near the point $(u,v) = (0,0)$. 
Let us consider the Taylor expansion of the function $g( r)$ at the first order
$$
g( r) = \frac{|r|}{2} + O(r^{2}).
$$
Observing that
\beq
\label{ineq}
\left | \frac{v}{2 \sqrt{u}}\right | \leq 1,
\eeq
as Riemann invariants are such that $\xi(x,t) \geq 0$ and $\eta(x,t) \leq 0$ for all $x$ and $t < t_{c}$, we have the following leading order approximation for the function $f(u,v)$
\begin{equation}
\label{fasym}
f(u,v) \simeq \frac{1}{24} \left(v^{2} + 8 u \right) \sqrt{4 u - v^{2}} + \frac{1}{2} u  v \sin^{-1} \left(\frac{v}{2 \sqrt{u}} \right).
\end{equation} 
The asymptotic expression~(\ref{fasym}) is obtained by splitting
the integral in Eq.~(\ref{fg}) as follows
\begin{eqnarray}
&\int_{-1}^{1}  \textup{sign}(v + 2 \mu \sqrt{u}) \sqrt{1-\mu^{2}} d \mu = \\
&- \int_{-1}^{-\frac{v}{2 \sqrt{u}}} \sqrt{1-\mu^{2}} d \mu  + \int_{-\frac{v}{2 \sqrt{u}}}^{1}\sqrt{1-\mu^{2}} d \mu, \nonumber
\end{eqnarray}
where the inequality~(\ref{ineq}) has been taken into account.
In terms of Riemann invariants the formula~(\ref{fasym}) reads as follows
\begin{eqnarray}
\label{FRasy}
F(\xi,\eta) \simeq & &\frac{1}{32} \left(\xi^{2} - \frac{2}{3} \eta \xi + \eta^{2} \right)  \left(- \eta \xi \right)^{1/2} +\nonumber \\
& &+\frac{1}{64} \left(\xi -\eta \right)^{2} (\xi + \eta) \sin^{-1} \left(\frac{\xi + \eta}{\xi -\eta} \right).
\end{eqnarray}
Hence, we can compute asymptotic expressions for the hodograph equations (\ref{hodR}) obtaining 
\begin{gather}
\label{hod_asym}
\begin{aligned}
x & \simeq  \frac{2 (-\eta \xi)^{3/2}}{(\xi - \eta)^{2}}, \\
t  & \simeq  - \frac{\xi + \eta}{(\xi-\eta)^{2}} (-\eta \xi)^{1/2} + \frac{1}{2} \sin^{-1} \left(- \frac{\xi+\eta}{\xi-\eta} \right),
\end{aligned}
\end{gather}
and also for the formulae~(\ref{rbreak}) 
\begin{equation}
\label{rbreakasy}
\xi_{x} \simeq  \frac{(\xi - \eta)^{2}}{2 \xi^{1/2} (-\eta)^{3/2}} \qquad
\eta_{x} \simeq   - \frac{(\xi - \eta)^{2}}{2 \xi^{3/2} (-\eta)^{1/2}}. 
\end{equation} 
From above expressions, it follows that the map $x =x(\xi,\eta)$, $t =t(\xi,\eta)$ is not invertible at $(\xi,\eta) = (0,0)$ as the limit for $(\xi,\eta)\to (0,0)$ does not exist in the usual sense. Such a limit depends on the specific path followed as a consequence of the fact that the vacuum point is preserved by the evolution. For instance, at a fixed time $t=t_{1}$ the configuration of Riemann invariants is associated to the path represented by the curve $(\xi_{1},\eta_{1}) = (\xi(x,t_{1}),\eta(x,t_{1}))$ parametrized by the independent variable $x$. Different times are associated to difference paths configurations and vice-versa. 
From the formulae~(\ref{rbreakasy}) we have that necessary condition for Riemann invariants to break is that at least one of the Riemann invariants has to be vanishing at the critical point, $\xi = 0$ or $\eta =0$. 
On the positive half-line $x>0$, the critical values such that $\eta_x$ blows-up and $\xi_x$ stays finite [see also Fig.~(\ref{SWEriemann}) for a visual representation] can be computed as the limit $(\xi,\eta) \to (0,0)$ along the  curve $\xi = 0$, i.e.
\begin{align*}
x_{c} &= \lim_{\eta \to 0} \left . x(\xi,\eta) \right |_{\xi = 0} = 0, \\
t_{c} &= \lim_{\eta \to 0} \left . t(\xi,\eta) \right |_{\xi = 0} = \lim_{\eta \to 0} \frac{1}{2} \sin^{-1} (1)  = \frac{\pi}{4} \simeq 0.785.
\end{align*}
Similarly, the limit along the curve $\eta = 0$ provides the following result  
\begin{align*}
x_{c} &= \lim_{\xi \to 0} \left . x(\xi,\eta) \right |_{\eta = 0} = 0, \\
t_{c} &= \lim_{\xi \to 0} \left . t(\xi,\eta) \right |_{\eta = 0} = \lim_{\eta \to 0} \frac{1}{2} \sin^{-1} (-1)  = - \frac{\pi}{4} \simeq  -0.785,
\end{align*}
that gives the critical values such that $\xi_x$ blows up on the positive $x$ semi-axis backward in time (equivalent, by symmetry, to negative half-line $x<0$, forward in time), and $\eta_{x}$ stays finite. From these limits we conclude that $t_c=\pi/4$ is the critical time at which both Riemann invariants break in the origin though in the limit from opposite semi-axes (i.e. $\xi_x$ and $\eta_x$ blow up in $x=0^-$ and $x=0^+$, respectively). This situation is summarized in Fig.~\ref{SWEriemann} which illustrates the essential feature of the breaking mechanism by showing snapshots of the behavior of the Riemann invariants $\xi(x)$ and $\eta(x)$ for different times.
As clearly shown, both Riemann invariants bend from the beginning, thereby increasing their slope in the origin (though on opposite semi-axis), until they experience a gradient catastrophe simultaneously at the origin as $t \rightarrow t_c=\pi/4$ \cite{nota_ux0}. 
Moreover, from asymptotic expressions~(\ref{rbreakasy}) clearly follows that the vacuum point is a non-generic breaking point as both denominators vanish.

\section{Dispersive effects}
\begin{figure}[h!] 
\label{figknee}
\centering
\includegraphics[width=6.5cm]{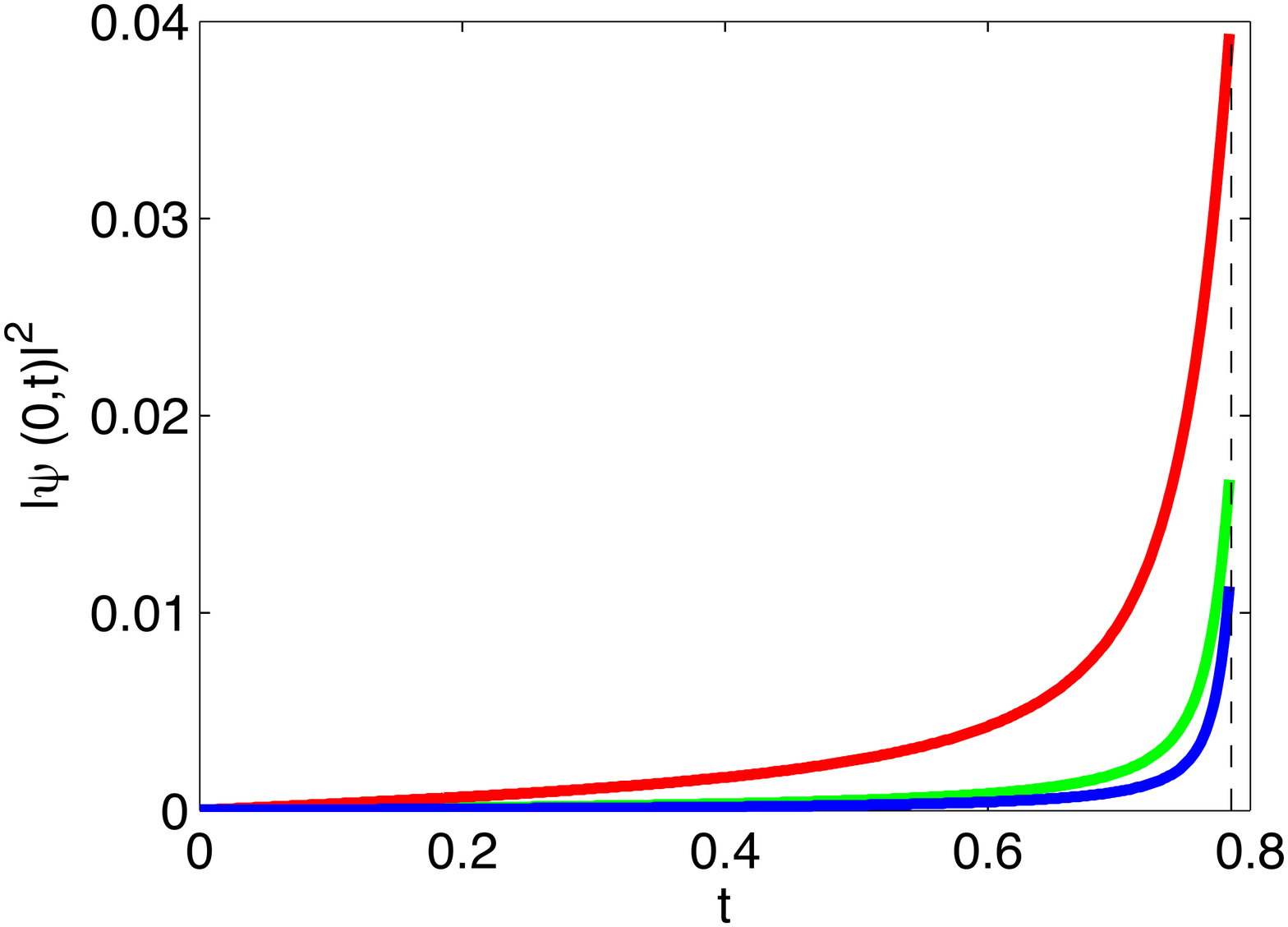}
\caption{(Color online) Density in the origin $u(0,t)=|\psi(0,t)|^2$ vs. time $t$, as obtained from numerical integration of NLS equation with decreasing values of dispersion: $\varepsilon=5 \times 10^{-3}$ (red curve), $\varepsilon=10^{-3}$ (green curve), $\varepsilon=5 \times 10^{-4}$ (blue curve). The dashed vertical line stands for the critical time $t_c=\pi/4$. Note the small values on the vertical axis.} 
\label{NLS1}\end{figure}

\begin{figure}[h!] 
\centering
\includegraphics[width=8cm]{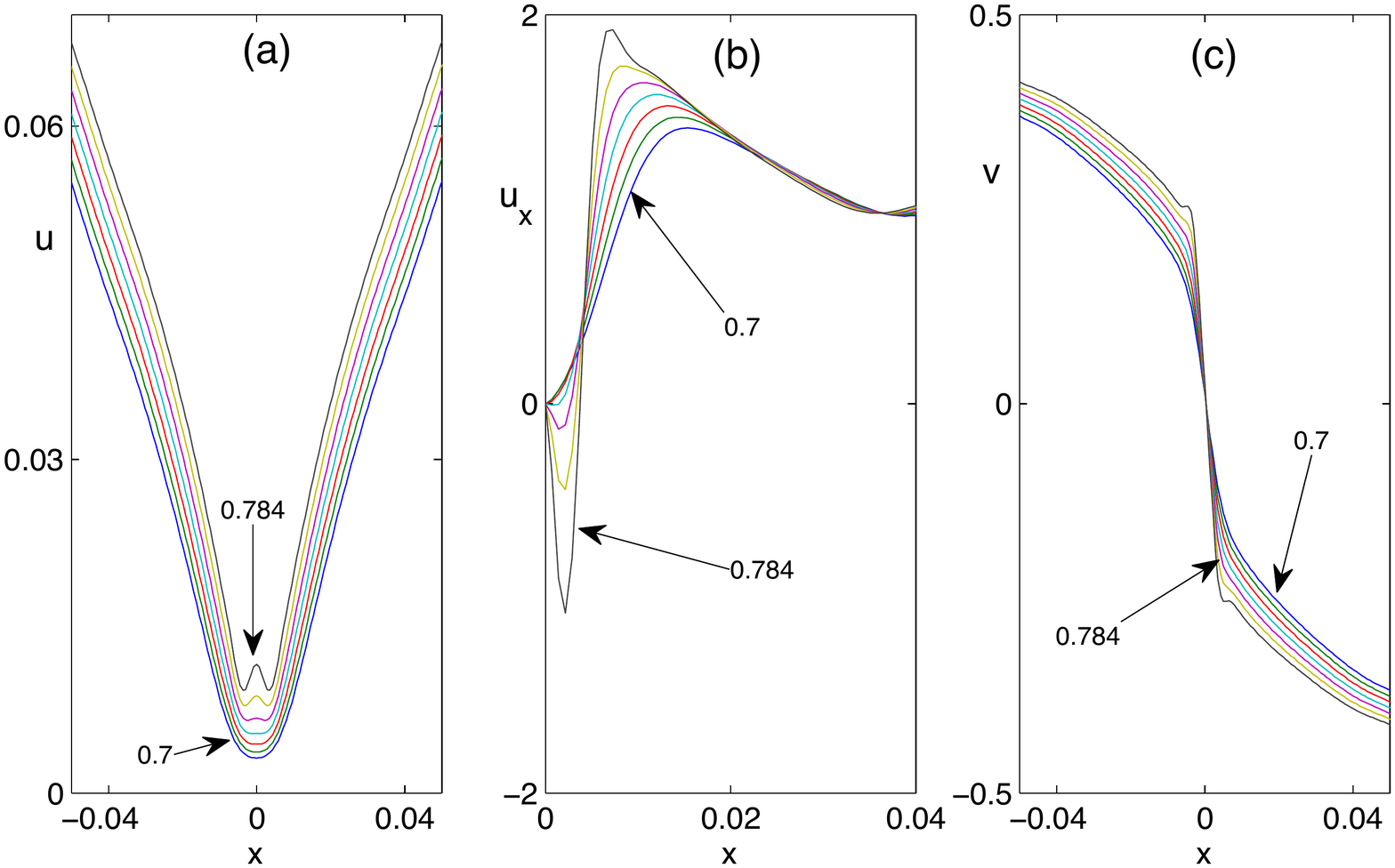}
\caption{(Color online) Hydrodynamical variables in the neighborhood of the origin, obtained from numerical integration of NLS equation with $\varepsilon=5 \times 10^{-4}$:
(a) $u$ vs. $x$;  (b) $u_x$ vs. $x$;  (c) $v$ vs. $x$. The snapshots are taken from time $t=0.76$ to $t=0.784$ with constant increment $\delta t=0.004$.} 
\label{NLS2}
\end{figure}

Let us study in more detail the behavior of the solution to the defocusing NLS equation~(\ref{NLS}) with the initial datum~(\ref{initdatum}). To this end we integrate numerically Eq. (\ref{NLS}) by means of a split-step algorithm with periodic boundary conditions in $x$, where the dispersive term is dealt with in Fourier space. The results, typically obtained with up to $2^{15}=32768$ equally spaced points over the window $\Delta x=12$ (spatial step $dx=3.7 \times 10^{-4}$) and time step $dt=5 \times10^{-6}$, are reported in Figs. \ref{NLS1} and  \ref{NLS2}. In particular, an important point to be emphasized is that, as a consequence of the dispersion, the vacuum point is no longer preserved by the evolution. Indeed, as shown in Fig.~(\ref{NLS1}), the density in the origin $u(0,t)=|\psi(0,t)|^2$ slowly detaches, starting at $t=0$, from the zero value characteristic of the SWE. However, this is not sufficient to say that the SWE fail to describe the underlying mechanism of breaking. First, we have verified that, far from the critical time, the deviation from the zero density value remains of order $O(\epsilon)$.  Even more importantly, as the critical time is approached, the density in the origin $u(0,t)$ starts to increase in a much faster way. In particular, runs performed with different values of $\varepsilon$ clearly show [see Fig.~(\ref{NLS1})] that this change of slope becomes more and more pronounced as $\varepsilon$ decreases. At $\varepsilon=5 \times 10^{-4}$, the curve $u(0,t)$ already exhibits a strong flattening towards zero before the critical time and a marked knee that is consistent with the expected behavior in the limit $\varepsilon \rightarrow 0$. 

The dynamics near the origin is further illustrated in Fig.~(\ref{NLS2}) which displays snapshots from the NLS equation in the neighborhood of $x=0$ at times close to the critical time. As illustrated, for $t \sim t_c$ a smooth, though quite abrupt transition occurs due to the density in the origin passing from a local minimum to a maximum, while its derivative $u_x$ in the origin remains zero. Correspondingly the velocity $v$ exhibits essentially the dynamics predicted by SWE limit, with the local slope in the origin increasing (though remaining finite due to the effect of non-zero dispersion) as the critical time is approached. 
The observed change of convexity must be ascribed to the effect of the weak (though finite) dispersion in connection with the flat initial phase.
This behavior turns out to be also consistent with the fact that beyond the critical time,  the dark (grey) solitons that compose the shock fan [see Fig. \ref{fig1}] appear in pairs with opposite velocities $v=\pm dx/dt$, with the symmetric solitons at the inner edges of the fan corresponding to the dips shown in Fig. \ref{NLS2}(a). Experimental evidence for this type of behavior was also given in Bose-Einstein condensates \cite{Dutton01}, while a similar phenomenon is known in optics \cite{twodarkexp} in the full dispersive case with $\epsilon=1$ where, instead of any dispersive shock wave, a dark (black) input with constant phase decays into a single pair of grey solitons with opposite velocities as also predicted by the inverse scattering approach to the NLS equation.
It is also interesting to compare the solution for a constant phase dark initial datum considered above with a dark initial datum of the form  $\psi(x,0)=\tanh x$  such that the phase has a jump at the origin \cite{CFPRT09}. In this case, as it results from the inverse scattering transform analysis \cite{Fratax08}, while the breaking mechanism outlined above remains valid, the initial vacuum point is preserved by the evolution even beyond the breaking time and it is accompanied by the formation of a dispersive shock wave with a central still ($v=0$, hence black) soliton. 
In this case no change of convexity is ever observed in the vacuum point. However, this is strictly related to the peculiar initial value characterized by the phase jump and is not generalizable (for instance  this feature is not preserved for generic phase profiles which are smooth and antisymmetric).

\begin{figure}[h!] 
\centering
\includegraphics[width=8.5cm]{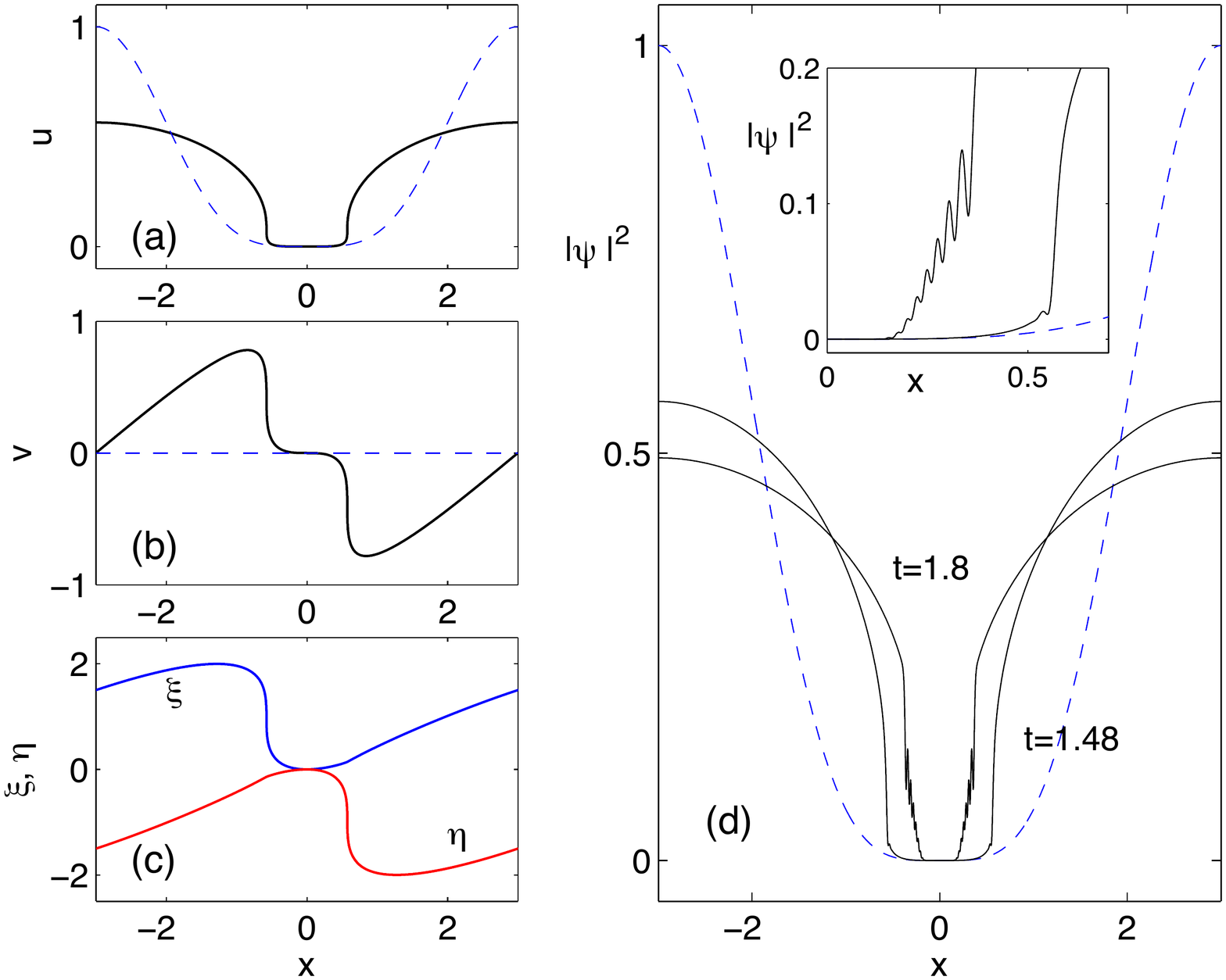}
\caption{(Color online) Evolution of initial waveform $\psi_0=[1- \cos(2\pi x/L)]/2$ (i.e., $u_0=[1- \cos(2\pi x/L)]^2/4$, $v_0=0$). Here $L=6$ is the numerical window.
(a-b-c) Snapshots of density $u$ (a), velocity $v$ (b), and Riemann invariants $\xi, \eta$ as obtained from SWE [Eqs. (6)] close to breaking ($t_c = 1.475$); 
(d) Snapshots of density $u=|\psi|^2$ obtained by integrating the NLS equation (1) with $\varepsilon=0.005$, right after breaking ($t=1.48$) when oscillations start to become visible, and at $t=1.8$ when they are fully developed. The inset shows details of the right dispersive shock (mirror symmetry in $x$ holds). In all panels, the blue dashed line is the input.} 
\label{raisedcos}\end{figure}

Finally we give an important remark concerning the generality of the mechanism of breaking investigated so far.
The mechanism of breaking that we have unveiled for the initial datum $\psi_0=|\tanh x|$ implies that breaking occurs exactly at the vacuum point (the origin). The latter is also the point responsible for the loss of strict hyperbolicity of the dispersionless system which is at the origin of such non-generic type of breaking. On this basis, one might conjecture that all (or at least most of) even initial data characterized by zero initial velocity and a single vacuum point in the density could follow a similar breaking dynamics. However it is not difficult to find counterexamples where even initial data exhibit generic breaking, yet exhibiting the same feature (i.e., strict hyperbolicity not holding due to a single vacuum point). In this case breaking does not occur in the vacuum point and involves a gradient catastrophe where both variables $u$ and $v$ exhibit diverging derivatives. This scenario generally occurs for input densities $u_0(x)=|\psi_0|^2$ that are sufficiently smooth around the vacuum point compared with the case $u_0(x)=\tanh^2 x$. Examples that we have verified include $\psi_0=\tanh^{2n}(x)$, for $n \ge 1$, or, in the periodic case, $\psi_0=[1- \cos(2\pi x/L)]/2$, with $L$ sufficiently large. As an example, we show the dynamics of the latter case (with $L=6$) in Fig. \ref{raisedcos}. In particular by numerically integrating the SWE, we find the occurrence of a generic breaking at the critical time $t_c \simeq 1.475$ [see Figs. \ref{raisedcos}(a,b,c)]. In this case the breaking points $x=\pm x_c$ are distinct from the vacuum ($x=0$) and reflect the invariance of the SWE under the inversion  symmetry $x,v \rightarrow -x, -v$. In this case the gradient catastrophe is accompanied by the divergence of the derivative of both $u$ and $v$ [Figs. \ref{raisedcos}(a,b)], while only one Riemann variable breaks at each point [Fig. \ref{raisedcos}(c)]. The integration of the NLS equation confirms this scenario, showing the build-up of oscillating wavetrains in the post shock evolution [see Fig. \ref{raisedcos}(d)]. This allows us to view the breaking mechanism of the initial datum $u_0(x)=\tanh^2 x$ as a degenerate case in which the breaking of the two Riemann invariants occurring in the generic case in two symmetric points, merge exactly in the vacuum point leading to a non-generic type of breaking. 
Results presented above show that the loss of strict hyperbolicity is not sufficient for the latter type of breaking to occur. 
On the other hand, the reader might wonder about the general requirements on the initial datum under which this transition occurs. 
This is a challenging problem that requires a different approach aimed at a general classification of the breaking for classes of initial data 
rather than the study of the solution to a specific initial value problem in the dispersionless limit. Although this remains clearly beyond the scope of the present paper, we believe that our results may be considered as a starting point for developing such an approach.
 
\section{Conclusions}
In summary, we have unveiled the breaking mechanism of a specific initial dark waveform with initial flat phase, which has its importance in relations to recent experiments performed in Bose-Einstein condensates and nonlinear optics and ruled by the defocusing NLS equation.
We have shown that the dispersionless equations undergo the simultaneous breaking of both Riemann invariants in the point of null density ($x=0$),
though breaking occurs in the opposite limits $x=0^{\pm}$ for the two invariants. The effect of dispersion starts to play a role before breaking and determines a slow adiabatic detachment of the minimum density points from zero. However, it is only around the breaking time predicted by the dispersionless limit that this minimum abruptly grows turning into a local maximum of the field. 
We believe that having enlightened such a mechanism, for the specific case considered here, represents a first step towards a classification of singularity types that can develop from initial data that contain vacuum points. This is however a challenging task to be faced in a future work.

\section*{Acknowledgements}
The authors wish to thank B. Dubrovin, P. Giavedoni and T. Grava for useful discussions and for providing references. A.M. has been partially supported by the ERC grant FroM-PDE and by SISSA Young Researchers Grant 2011 ``Nonlinear Asymptotics for Nonlinear Optics". S.T. acknowledges funding from MIUR (grants PRIN  2009P3K72Z and 2012BFNWZ2).

\appendix
\section{}
Here we show how we evaluate elliptic integrals appearing in Eq. (\ref{FR}):
\begin{widetext}
\label{app1}
\begin{align*}
\frac{1}{4} \int_{0}^{\xi} \frac{r \sqrt{(\xi-r) (r-\eta)}}{\sqrt{4-r^{2}}} \; dr=&  - 2 \int_{0}^{\xi} \sqrt{\frac{(\xi-r)(r-\eta)}{(2-r)(2+r)}} \; dr + \int_{0}^{\xi} \sqrt{\frac{(\xi-r) (r-\eta) (2+r)}{(2-r)}} \; dr \\
=&-2 \left [ \frac{\gamma}{2 \alpha^{2} (k^{2} -\alpha^{2})} (2-\xi) (\xi-\eta) \left(\alpha^{2} E(\omega) + (\alpha^{2} -k^{2}) \omega \right . \right . \\ 
&+ \left . \left . (\alpha^{4} -2 \alpha^{2} + k^{2}) \Pi(\omega,\alpha^{2}) - \alpha^{4} \frac{\textup{sn}\omega \;  \textup{cn} \omega \; \textup{dn}\omega}{1-\alpha^{2} \textup{sn}^{2}\omega}  \right) \right] \\
&+ \frac{\gamma}{\alpha^{4}} (2-\xi) (2+\xi) (\xi-\eta) \left [-k^{2} \omega + (3 k^{2} - \alpha^{2} k^{2} - \alpha^{2}) \Pi(\omega,\alpha^{2}) \right .  \\
& \left .+ (2 \alpha^{2} k^{2} + 2 \alpha^{2} - 3 k^{2} - \alpha^{4}) V_{2} + (\alpha^{2}-1) (\alpha^{2}-k^{2}) V_{3} \right ]
\end{align*}
with the notations
\begin{align*}
V_{0} =& {\cal F}(\phi,k)  \qquad  V_{1} = \Pi(\phi,\alpha^{2}, k) \\
V_{2} =&\frac{1}{2 (\alpha^{2} - 1) (k^{2} - \alpha^{2})} \left [\alpha^{2} E(\omega) +(k^{2} - \alpha^{2}) \omega + (2 \alpha^{2} k^{2} + 2 \alpha^{2} - \alpha^{4} - 3 k^{2}) \Pi(\phi,\alpha^{2},k) \right . \\  
&\left .- \alpha^{4} \frac{\textup{sn}\omega \;  \textup{cn}\omega \; \textup{dn}\omega}{1-\alpha^{2} \textup{sn}^{2}\omega}  \right ] \\
V_{3} =& \frac{1}{4 (1-\alpha^{2}) (k^{2} - \alpha^{2})} \left [k^{2} V_{0} + 2 (\alpha^{2} k^{2} + \alpha^{2} - 3 k^{2}) V_{1} + 3 (\alpha^{4} -2 \alpha^{2} k^{2} - 2 \alpha^{2} + 3 k^{2}) V_{2}   
\right .  \\ 
&\left . + \alpha^{4}  \frac{\textup{sn}\omega \;  \textup{cn}\omega \; \textup{dn}\omega}{(1-\alpha^{2} \textup{sn}^{2}\omega)^{2}}  \right ]
\end{align*}
\begin{align*}
&\alpha^{2} = \frac{\xi -\eta}{2-\eta} \qquad k^{2} =\frac{4 (\xi - \eta)}{(2-\eta) (2+\xi)} \qquad \gamma =\frac{2}{\sqrt{(2-\eta) (2+\xi)}} \\
& \phi =\sin^{-1} \left(\frac{\xi (2-\eta)}{2 (\xi-\eta)} \right )  \qquad \omega = \textup{sn}^{-1} (\phi,k^{2}) 
\end{align*}
where ${\cal F}(\phi,k)$, $E(\omega)$ and $\Pi(\phi,\alpha^{2},k)$ are the standard notations for the elliptic integral of first kind, the complete elliptic integral of second kind and the incomplete elliptic integral of the third kind  respectively and $\textup{sn}$, $\textup{cn}$, $\textup{dn}$ stand for the Jacobian Elliptic functions.
A similar formula for the second integral in~(\ref{FR}) is simply obtained from the above via the substitution
$(\xi,\eta) \to (-\eta,-\xi)$.

\end{widetext}


\begin{thebibliography}{99}

\bibitem{CAL}
F. Calogero, \emph{Why are certain nonlinear PDEs both widely applicable and integrable}, in ``What is integrability", editor V.E. Zhakarov, (Springer, Berlin, 1991), pp. 1--62.

\bibitem{agrawalbook}
Y. S. Kivshar and G.P. Agrawal, {\it Optical solitons: from fibers to photonic crystals} (Academic Press, San Diego, 2003);

\bibitem{osbornebook} A. Osborne, {\em Nonlinear Ocean Waves an the Inverse Scattering Transform} (Academic Press, New York, 2010).

\bibitem{GPeq} V. L. Ginzburg, L. P. Pitaevskii, Zh. Eksp.Teor. Fiz. {\bf 34}, 1240 (1958); E. P. Gross, 
J. Math. Phys. {\bf 4}, 195 (1963).

\bibitem{RB} P. H. Roberts and N. G. Berloff, {\em The Nonlinear Schr\"odinger Equation as a Model of Superfluidity},
in "Quantized Vortex Dynamics and Superfluid Turbulence" edited by C.F. Barenghi, R.J. Donnelly and W.F. Vinen, 
Lecture Notes in Physics, vol. 571, (Springer, 2001).

\bibitem{DGPS99} F. Dalfovo, S. Giorgini, L. Pitaevskii, and S. Stringari, Rev. Mod. Phys. {\bf 71}, 463 (1999).

\bibitem{CC13} 
I. Carusotto and C. Ciuchi, 
Rev. Mod. Phys. {\bf 85}, 299 (2013).

\bibitem{AS} M. Ablowitz and H. Segur, {\it Solitons and Inverse Scattering Transform}, SIAM, Philadelphia (1981). 

\bibitem{NMPZ} 
S.P. Novikov, S.V. Manakov, L.P. Pitaevskii and V.E. Zakharov {\it  Theory of solitons: The inverse scattering method}, Consultants Bureau, New York (1984).

\bibitem{ZS} 
V.E. Zakharov and P.B. Shabat, 
Sov. Phys. JEPT {\bf 34}, 62 (1972).

\bibitem{APT} M. Ablowitz, B. Prinari, D. Trubach, {\it Discrete and Continuous Nonlinear Schr\"odinger Systems},  Cambridge University Press (Cambridge, 2004).

\bibitem{BBEIM} E.D. Belokolos, A.I. Bobenko, V.Z.  EnolÕski, A.R. Its, V.B. Matveev, 
Springer Series in Nonlinear Dynamics, (Springer, Berlin, 1994).

\bibitem{ZK65} 
N. J. Zabusky and M.D. Kruskal, 
Phys. Rev. Lett. {\bf 15}, 240 (1965).

\bibitem{GP} V. Gurevich and L. P. Pitaevskii, 
Sov. Phys. JETP, {\bf 38}, 291 (1974).

\bibitem{UB} 
T.B. Benjamin and M.J. Lighthill, 
Proc. Roy. Soc. Lond. A, {\bf 224}, 448 (1954);
D. H. Peregrine,  
J. Fluid Mech. {\bf 25}, 321 (1966).

\bibitem{Whitham65}
G.B. Whitham, Proc. Roy. Soc. {\bf 283},  A238 (1965).

\bibitem{FFM80} 
H. Flaschka, M. G. Forest, and D. W. McLaughlin, 
Comm. Pure Appl. Math. {\bf 33}, 739 (1980).

\bibitem{LL83} 
P.D. Lax, C.D. Levermore, 
Comm. Pure Appl. Math. {\bf 36}, 253; {\bf 36}, 571; {\bf 36}, 809 (1983).

\bibitem{V85} 
S. Venakides, 
Comm. Pure Appl. Math. {\bf 38}, 833 (1985).

\bibitem{K} 
I. M. Krichever, 
Funct. Anal. Appl. {\bf 22}, 37 (1988).

\bibitem{Tian94} 
F.-R. Tian,  
Commun. Math. Phys.  {\bf 166}, 79 (1994).

\bibitem{GK87}
A.V. Gurevich and A. L. Krylov, 
Sov. Phys. JETP {\bf 65}, 944 (1987).

\bibitem{P87}  
M. V. Pavlov, 
Teoret. Mat. Fiz. {\bf 71}, 351 (1987), in Russian; Theoret. Math. Phys., {\bf 71}, 584 (1987), English translation.


\bibitem{GKE92}
A.V. Gurevich, A. L. Krylov, and G.A. El,
Sov. Phys. JETP {\bf 74}, 957 (1992).

\bibitem{ElKrylov95}
G.A. El and A. L. Krylov, 
Phys. Lett. A {\bf 203}, 77 (1995).

\bibitem{EGGK95} 
G. A. El, V. V. Geogjaev, A. V. Gurevich, and A. L. Krylov, 
Physica D {\bf 87}, 186 (1995).


\bibitem{TY99} 
F.-R. Tian and J. Ye,  
Comm. Pure Applied Math. {\bf 52}, 655 (1999).

\bibitem{JLM1} S. Jin, C.D. Levermore, D.W. McLaughlin, {\it The behavior of solutions of the NLS equation in the semiclassical limit. Singular limits of dispersive waves}, pp. 235-255, N. M. Ercolani, I. R. Gabitov, C. D. Levermore, and D. Serre, Eds., NATO Adv. Sci. Inst. Ser. B Phys., vol. 320 (Plenum, New York, 1994).

\bibitem{JLM2} S. Jin, C. D. Levermore, and D. W. McLaughlin, 
Comm. Pure Applied Math. {\bf 52}, 613 (1999).

\bibitem{WFM99}
O. C. Wright, M. G. Forest, and K. T.-R. McLaughlin, 
Phys. Lett. A {\bf 257}, 170 (1999).

\bibitem{Kodama99} 
Y. Kodama, 
SIAM J. Appl. Math. {\bf 59}(6), 2162 (1999).

\bibitem{Kam04}
A. M. Kamchatnov, A. Gammal, and R.A. Kraenkel, 
Phys. Rev. A {\bf 69}, 063605 (2004). 

\bibitem{Hoefer06} M.A. Hoefer, M.J. Ablowitz, I. Coddington, E.A. Cornell, P. Engels, and V. Schweikhard, 
Phys. Rev. A {\bf 74}, 023623 (2006).

\bibitem{Fratax08}
A. Fratalocchi, C. Conti, G. Ruocco, and S. Trillo, 
Phys. Rev. Lett. {\bf 101}, 044101 (2008).


\bibitem{Dutton01}
Z. Dutton, M. Budde, C. Slowe, L. V. Hau, 
Science {\bf 293}, 663 (2001).

\bibitem{Chang08}
J. J. Chang, P. Engels, and M. A. Hoefer, 
Phys. Rev. Lett. {\bf 101}, 170404 (2008).
 

\bibitem{Mep09} R. Meppelink, S. B. Koller, J. M. Vogels, P. van der Straten, E. D. van Ooijen, N. R. Heckenberg, H. Rubinsztein-Dunlop, S. A. Haine, M. J. Davis, 
Phys. Rev. A  {\bf 80}, 043606 (2009).

\bibitem{RG89}
J. E. Rothenberg and D. Grischkowsky,  
Phys. Rev. Lett. {\bf 62}, 531 (1989).

\bibitem{Wan07}
W. Wan, S. Jia, And J. W. Fleischer, 
Nature Phys. {\bf 3}, 46 (2007).

\bibitem{Gofra07} 
N. Ghofraniha, C. Conti, G. Ruocco, S. Trillo, 
Phys. Rev. Lett. {\bf 99}, 043903 (2007).

\bibitem{Fleischer07} 
S. Jia, W. Wan, and J. W. Fleischer, 
Phys. Rev. Lett. {\bf 99}, 223901 (2007).

\bibitem{CFPRT09} C. Conti, A. Fratalocchi, M. Peccianti, G. Ruocco and S. Trillo, 
Phys. Rev. Lett. {\bf 102}, 083902 (2009).

\bibitem{D06} B. Dubrovin, 
Comm. Math. Phys.  {\bf 267}, 117, (2006).

\bibitem{D08} B. Dubrovin,  
Amer. Math. Transl. Ser. 2, {\bf 224}, 59 (2008).

\bibitem{CG09} T. Claeys and T. Grava, 
Commun. Math. Phys. {\bf 286}, 979 (2009).

\bibitem{DGK09} 
B. Dubrovin, T. Grava, C. Klein, 
J. Nonlinear Sci. {\bf 19}, 57 (2009).

\bibitem{BT12} M. Bertola and A. Tovbis, 
Comm. Pure Appl. Math. {\bf 66}, 678 (2013).

\bibitem{DGKM} 
B. Dubrovin, T. Grava, C. Klein, A. Moro, 
arXiv:1311.7166v1.

\bibitem{DE} B. Dubrovin and M. Elaeva, 
Russian J. Math. Phys. {\bf 19}, 13 (2012).

\bibitem{ALM} A. Arsie, P. Lorenzoni, A. Moro, 
{\tt arXiv:1301.0950v2} (2013).

\bibitem{Armaroli09}
A. Armaroli, S. Trillo, A. Fratalocchi,
Phys. Rev. A {\bf 80}, 053803 (2009).

\bibitem{Trillo10}
S. Trillo and A. Valiani,
Opt. Lett. {\bf 35}, 3967 (2010).

\bibitem{Smoller80} 
T.-P. Liu and  J. A. Smoller, 
Adv. Appied Math. {\bf 1}, 345 (1980).

\bibitem{Yang06} 
T. Yang, 
J. Comp. Applied Math. {\bf 190}, 211 (2006).

\bibitem{ER70} 
A. Erd\'elyi,  
J. d' Analyse Math\'{e}matique {\bf 23}, 89 (1970).

\bibitem{twodarkexp}
D. Kr\"okel, N.J. Halas, G. Giuliani, and D. Grischkowsky, 
Phys. Rev. Lett. {\bf 60}, 29 (1988); A. M. Weiner, J. P. Heritage, R. J. Hawkins, R. N. Thurston, E. M. Kirschner, D. E. Leaird, W. J. Tomlinson, 
Phys. Rev. Lett. {\bf 61}, 2445 (1988); Y. S. Kivshar and B. Luther-Davies, 
Phys. Rep. {\bf 298}, 81-197 (1998).

\bibitem{BF} P.F. Byrd and M.D. Friedmann, {\it Handbook of Elliptic Integrals for Engineers and Scientists}, Springer-Verlag, Berlin (1971).

\bibitem{nota_ux0} Since $t=0.78 < t_c$, the inset of Fig. 3 could erroneously suggest that the derivative $u_x(0) \neq 0$, contradicting our previous statement. 
However, a closer look over a shorter range in $x$ shows indeed that $u$ approaches the zero value at $x=0$ with vanishing derivative.


\end{thebibliography}
\end{document}